# Unusual Ferroelectricity of Trans-Unitcell Ion-Displacement and Multiferroic Soliton in Sodium and Potassium Hydroxides

Yangyang Ren[1], Shuai Dong[2] and Menghao Wu[1]*

[1]School of Physics, Huazhong University of Science and Technology, Wuhan, Hubei, China

[2]School of Physics, Southeast University, Nanjing, Jiangsu, China

*Email: wmh1987@hust.edu.cn

Abstract

We show the first-principles evidence of a hitherto unreported type of ferroelectricity with ultra-long ion-displacement in sodium and potassium hydroxides. Even a small amount of proton vacancies can completely change the mode of proton-transfer from intra-unitcell to trans-unitcell, giving rise to multiferroic soliton with "mobile" magnetism and a tremendous polarization that can be orders of magnitude higher compared with most perovskite ferroelectrics. Their vertical polarizations of thin-film are robust against depolarizing field, rendering various designs of 2D ferroelectric field-transistors with non-destructive readout and ultra-high on/off ratio via sensing the switchable metallic/insulating state.

**Introduction**

Ferroelectric (FE) materials are polar substances with spontaneous electric polarizations that can be switched by an external electric field, which are indispensable for electronics, micromechatronics, and electro-optics. Traditional ferroelectrics like pervoskites[1] have already found a variety of commercial uses, such as nonvolatile memories[2], piezoelectric actuators, thermal sensors, light modulators, etc., but still often include toxic lead or rare metals such as bismuth, niobium or tantalum as the component elements. Moreover, in their ultrathin films, FE perpendicular to the film surface is suppressed by the depolarizing field and will disappear below film thicknesses (like 24 Å in $BaTiO_3$, 12 Å in $PbTiO_3$)[3-4]. After all, the displacements of metal ions during FE switching are within 0.5 Å and can be diminished by depolarizing field. To resolve this issue for the high-density integration of non-volatile memories, a variety of new FE systems have been explored recently, including FE based on emerging types of two-dimensional (2D) van der Waals materials as summarized in our recent review.[5]

In this paper we will focus on prototypical strong base with industrial and niche applications: sodium and potassium hydroxides, which are partially ionic and partially hydrogen-bonded solids as the hydroxyl groups OH− form hydrogen-bonded networks. However, it has not yet been noticed that FE may be induced by proton-transfer in the hydrogen-bonded network of NaOH and KOH. In such proton-transfer FE[6-10] where polarity can be formed spontaneously for the direction preference of hydrogen bonding even in 1D,[6] steric hindrance or high energy barriers during switching can be avoided and strong hydrogen bonds may give rise to high Curie temperature. Through first-principles calculations we predict the existence of robust proton-transfer FE with considerable polarization in NaOH and KOH, which enable mass production compared with traditional FE. Moreover, a new mode of FE switching with trans-unitcell ion-displacement may emerge upon a small amount of proton vacancies, giving rise to a tremendous polarization and multiferroic soliton with mobile magnetic moment. Their thin-film possess vertical polarizations that are robust against depolarizing field, which can be used as FE field-transistor for nonvolatile control of on/off state for 2D materials.

**Methods**

The theoretical calculations are performed based on density functional theory (DFT) methods implemented in the Vienna Ab initio Simulation Package (VASP 5.3.3) code.[11-12] The generalized gradient approximation (GGA) in the Perdew−Burke−Ernzerhof (PBE)[13] form for the exchange and correlation potential, together with the projector-augmented wave (PAW)[14] method, are adopted. In particular, PBE-D2 functional of Grimme[15] is used to account for the van der Waals interaction. The kinetic energy cutoff is set to be 520 eV, and computed forces on all atoms are less than 0.001 eV/Å after the geometry optimization. For electronic band-structure computation, the Brillouin zone is sampled using 2×6×6 and 4×8×4 k-point grid in the Monkhorst-pack scheme.[16] The Berry phase method is employed to evaluate crystalline polarization,[17] and the FE switching pathway is obtained by using nudged elastic band (NEB) method.[18] The phonon band structure was calculated using density functional perturbation theory (DFPT).[19] All performed DFPT calculations were started from the fully relaxed structures.

**Results and Discussion**

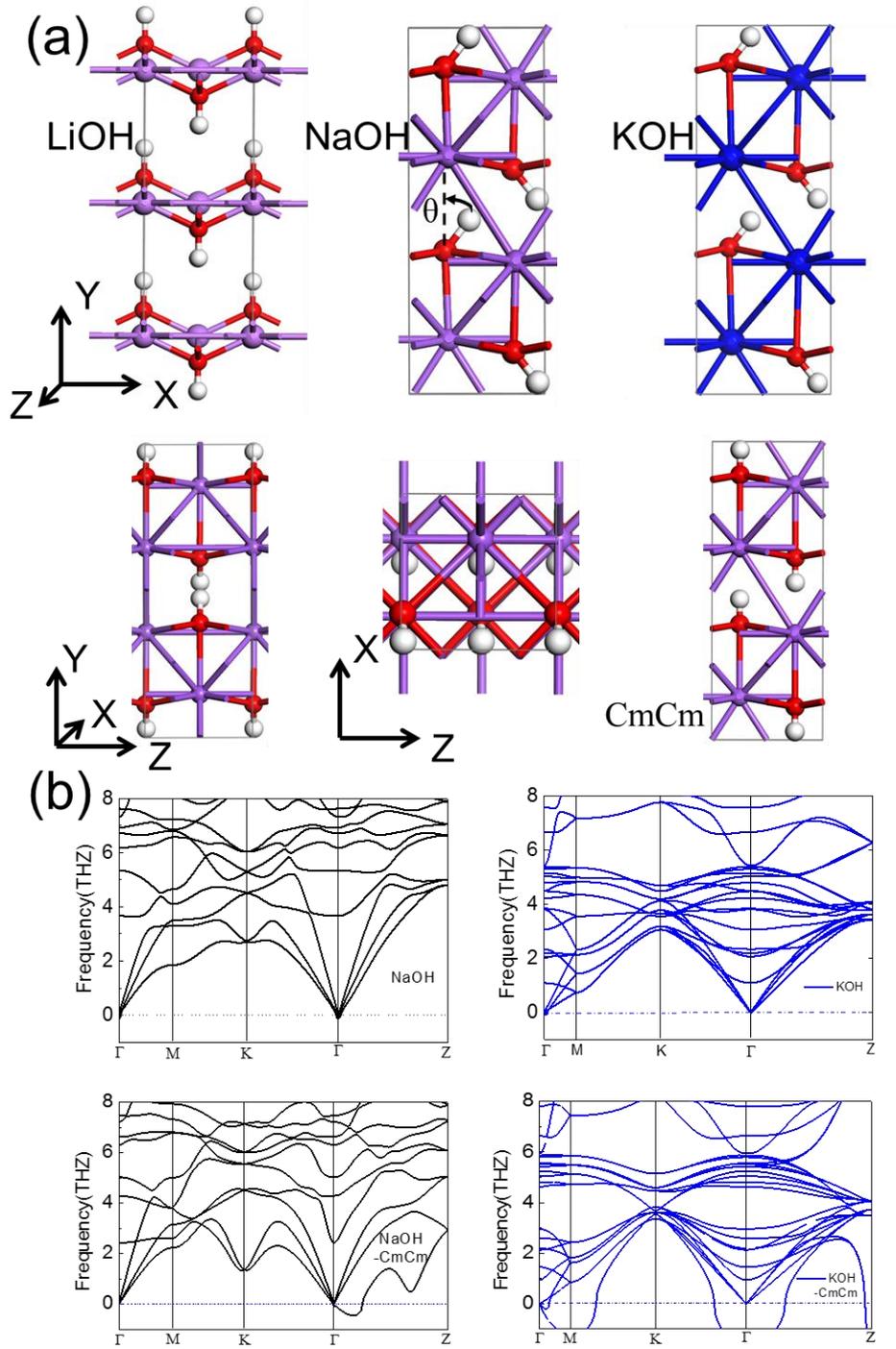

Fig. 1 (a)Geometric structure of LiOH, NaOH and KOH, where pink, purple, blue, red, white spheres denote Li, Na, K, O, H atoms respectively.(b) Phonon dispersions of NaOH and KOH for polar (ground ) and nonpolar (CmCm) state.

The optimized geometric structures of LiOH, NaOH and KOH are displayed in Fig. 1(a), based on previous infrared spectra and proton nuclear magnetic resonance studies[20-21]. Here the structure of LiOH possesses a highly symmetrical group of P4/nmm where the –OH bonds are aligned parallel along –z direction, while for the ground state of NaOH and KOH, the direction of –OH bonds deviate from the –y axis by angle θ= 39.3 and 43.6 degree, respectively. As a result, polar Ccm21(C2V-12) structures are formed which are respectively 0.075 and 0.14eV/f.u. lower than the symmetrical CmCm state where the –OH bonds are aligned parallel along –z direction. Due to the proton deviation, all the –OH groups form into bundles of hydrogen-bonded chains, where the bond length of hydrogen bond O…H in NaOH and KOH are respectively 1.76 and 1.86 Å. The phonon spectrum of polar ground state and non-polar phase for NaOH and KOH are plotted in Fig. 1(b), where the imaginary soft optical modes of the non-polar phase indicate the spontaneous symmetry-breaking below Curie temperature. Meanwhile, the dynamical stability of polar ground state without imaginary frequency in phonon dispersion is further verified for both NaOH and KOH.

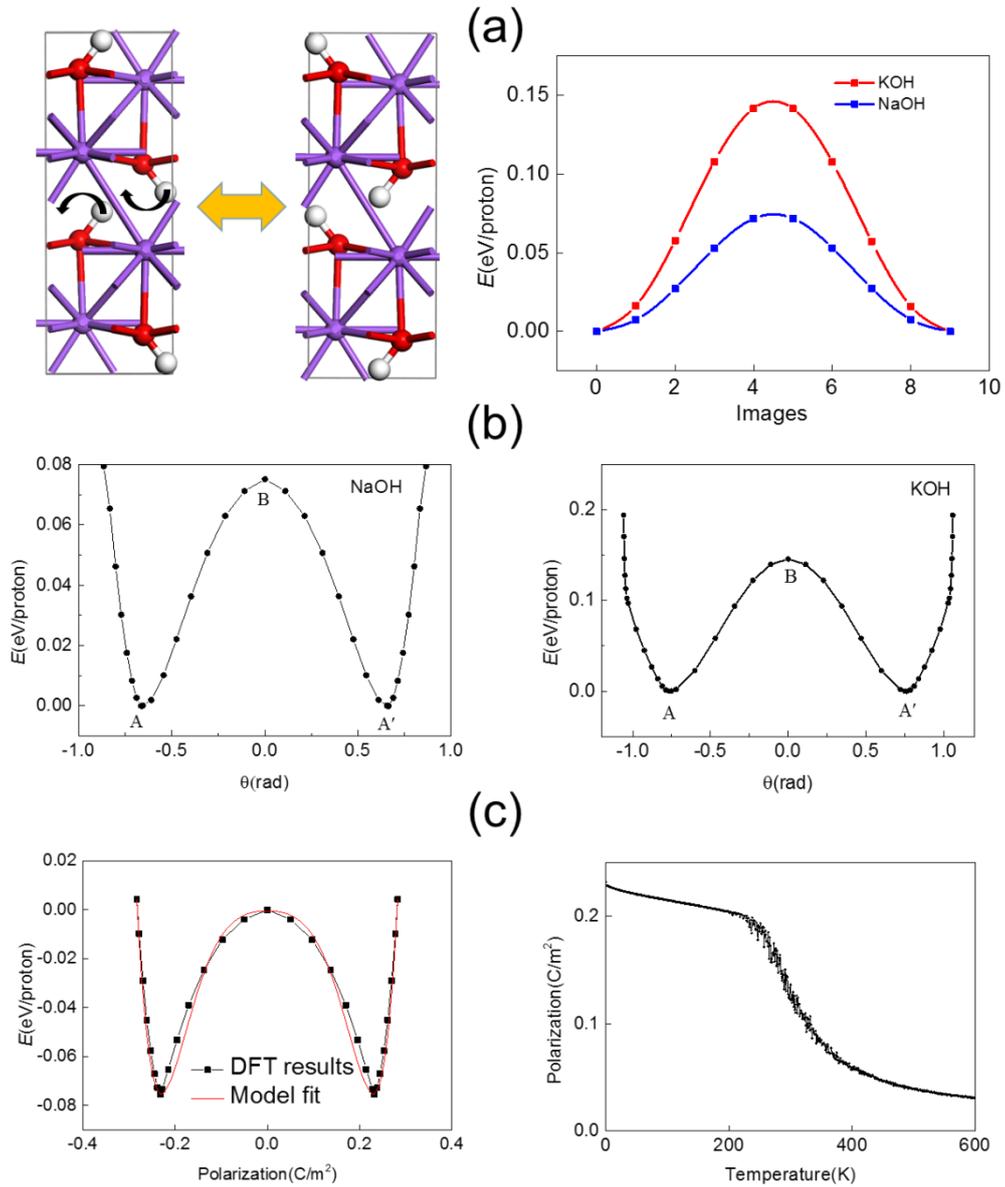

Fig. 2 (a) The FE switching pathway upon swirling of -OH bonds. (b) Double well potential of NaOH and KOH vs the deviation angle θ of -OH. (c) Double well potential of NaOH vs the polarization, and temperature dependence of polarization obtained from Monte Carlo simulations.

The symmetry breaking induced by proton deviations in NaOH and KOH may give rise to FE, as long as the polarization is switchable and the system is insulating. To our calculations, the bandgap of NaOH and KOH are both larger than 6eV (see Fig. S1). If the angle between –OH and z axis is able to swirl from -$\theta_{max}$ to $\theta_{max}$, the switchable bulk polarization of NaOH and KOH will be respectively 23.2 and 17.6 μC/cm$^2$, comparable to perovskite FE like BaTiO$_3$. The energy profiles of FE switching pathway as a function of step number are plotted in Fig. 2(a) using NEB method, with the angle between –OH and z axis swirling from -$\theta_{max}$ to $\theta_{max}$. The obtained switching barrier for NaOH and KOH are respectively 0.075 and 0.14eV/f.u., which may be even lower if the nuclear quantum effect of protons is taken into account. The existence of FE in NaOH and KOH can be further verified by the anharmonic double-well potential vs the deviation angle of -OH in Fig. 2(b), where the barrier of the double-well potential is consistent with the results obtained by NEB calculations.

Alternatively, those systems can be described by the Landau theory. Here NaOH is selected as the example for study, where the potential energy is expressed in the Landau-Ginzburg expansion

$$E = \sum_i \frac{A}{2}(P_i^2) + \frac{B}{4}(P_i^4) + \frac{C}{6}(P_i^6) + \frac{D}{2}\sum_{\langle i,j \rangle}(P_i - P_j)^2 \tag{1}$$

which can be viewed as the Taylor series of local structural distortions with a certain polarization defined at each cell $P_i$. As shown in Fig. 2(c), the first three terms are associated with the energy contribution from the local modes up to the sixth order, which can well describe the anharmonic double-well potential. The last term captures the coupling between the nearest local modes, which can be obtained by mean-field theory within the nearest-neighbor interaction. The fitted parameters A, B, C, D turn out to be respectively -0.5466, 288.58, 5718.85 and 0.9322. With the effective Hamiltonian and parameters, Monte Carlo simulation is used to investigate the phase transition. Its Curie temperature over 300K is obtained, which accords with our results of molecular dynamics simulation at 300K shown in Fig. S2.

Besides the swirling of -OH bonds, the hopping of protons may induce another type of FE. As shown in Fig. 3(a), if the protons are able to simultaneously hop to adjacent O atoms along the hydrogen bonds, the switchable polarization will be 26.1 and 22.6 $\mu C/cm^2$ respectively for NaOH and KOH. To our NEB calculation, the hopping barrier are respectively 0.23 and 0.28 eV, larger than the -OH swirling barrier. Since all the –OH groups form into bundles of quasi-one-dimensional (1D) hydrogen-bonded chains in NaOH and KOH crystal, a model of 1D hydrogen-bonded chain is built in Fig. 3(b). Herein the proton hopping may be prohibited in perfect crystal but accessible with proton vacancy: for a proton vacancy hopping to its adjacent hydrogen-bonded O atom, the energy barrier is 0.12 and 0.15 eV respectively in NaOH and KOH, much lower than the simultaneously hopping barrier of protons without vacancy. It is known that proton transfer in a hydrogen-bonded chain subjected to an appropriate electrostatic field may behave as a soliton.[22-23] Upon an electric field, the vacancy will be driven to one end of chain, and stop at the interface between electrodes and NaOH/KOH if the electrodes are proton-insulating; similarly, it will be driven to the other end when the electric field is reversed, as shown by the successive trans-unitcell proton-transfer mode in Fig. 3(c). Here the bulk property is completely changed by just one vacancy, as the displacement of all protons upon FE switching will be prolonged by half of a periodic length along the chain. The switchable polarization will be respectively 72.5 and 57.8 $\mu C/cm^2$ with one proton vacancy for each hydrogen-bonded chain in NaOH and KOH. If there are N vacancies along each chain, theoretically the transfer distance of each proton upon FE switching will be further greatly enhanced, and the switchable polarization will be respectively 23.2+49.3N and 17.6+40.2N µC/cm$^2$ for NaOH and KOH, which can be inconceivable values even to unlimited large: even if only 0.1% protons are missing in a cubic NaOH crystal with size ~1 µm, N=~6 and the polarization of NaOH$_{0.999}$ will be ~300 µC/cm$^2$. Moreover, each proton vacancy induces a magnetic moment of 1µ$_B$ mainly distributed around the O atom with no proton bonded around (see the spin density distribution around proton vacancy in Fig. 4(d)), which will follow the migration of vacancy along the hydrogen-bonded chain. As a result, the soliton will be multiferroic with "mobile" magnetism along the chain, which also renders a coupling between FE and magnetism for efficient data reading and writing.

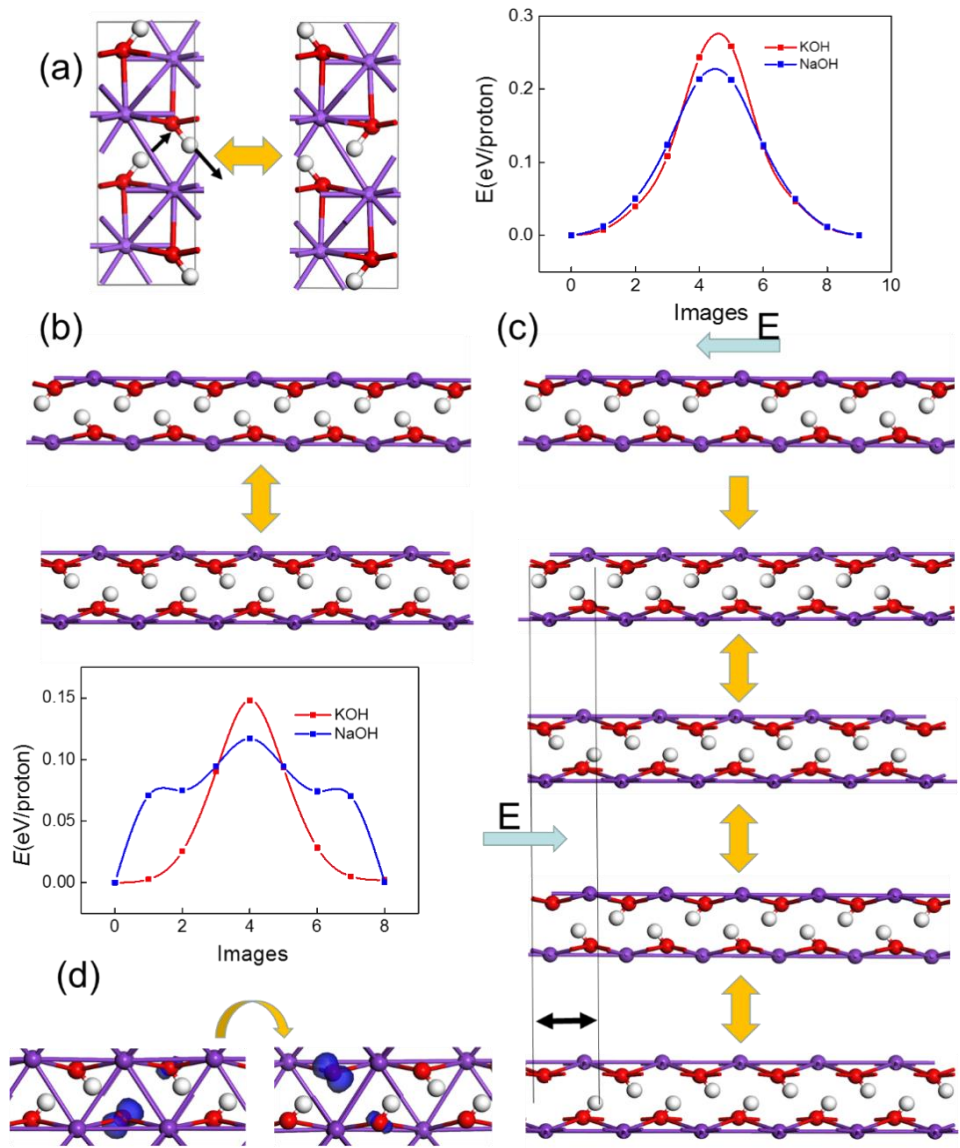

Fig. 3 (a)The FE switching pathway by proton hopping in NaOH and KOH. (b) Model of proton-transfer along 1D hydrogen-bonded chain without vacancy. (c) Model of proton-transfer along 1D hydrogen-bonded chain with one proton vacancy and its FE switching pathway.(d) Hopping pathway of a proton vacancy, where the spin density distribution is plotted in blue.

It is known that for ultrathin films of traditional FE, the vertical polarization is suppressed by the depolarizing field and will disappear or switch in-plane below critical thicknesses.

However, for both NaOH and KOH the polarization is aligned along the direction of hydrogen-bonded chains, which can be hardly diminished or turned to other direction. If the chain direction is set as the vertical direction for a 6-layer KOH thin-film, the polarization will be $0.3\times10^{-10}$C/m, much larger than the vertical polarization obtained in previously predicted 2D FE (e.g., $0.02\times10^{-10}$C/m for bilayer BN[24], $0.11\times10^{-10}$C/m for intercalated phosphorene bilayer[25]) . The polarization will be switchable at ambient condition as the switching barrier in thin-film is lowered to 0.06eV/proton (see the structure and pathway in Fig.4(a) ). Moreover, the polarization can be even greatly enhanced with a small amount of proton vacancies, which also induce electrical-tunable magnetism. It can be used as the FE dielectrics on 2D materials for non-volatile control of field-effect transistor,[26] as shown in Fig. 4(b): electron accumulates as the FE polarization is directed toward 2D materials, while depletes as the polarization is directed opposite. Since NaOH and KOH are uniaxial FE along the chain direction, the in-plane polarization can also be switched upon vertical FE switching if the chain orientation is set with a tilted angle from vertical direction. Herein 2D InSe and $Bi_2Se_3$ monolayer that have been experimentally verified to be semiconductors with high electron mobility[27] are chosen as two examples for design, also due to their perfect lattice match with KOH for DFT calculations. As shown in Fig. 4(c) and (d), both systems of KOH/InSe and KOH/$Bi_2Se_3$ heterostructures are semiconducting with moderate bandgaps when the FE polarizations of KOH point upwards, while both become metallic when the FE polarizations switch downwards and towards the 2D materials.[28] The switchable semiconducting/metallic states render an ultra-high on/off ratio for non-destructive readout via sensing two distinct source-drain resistances.

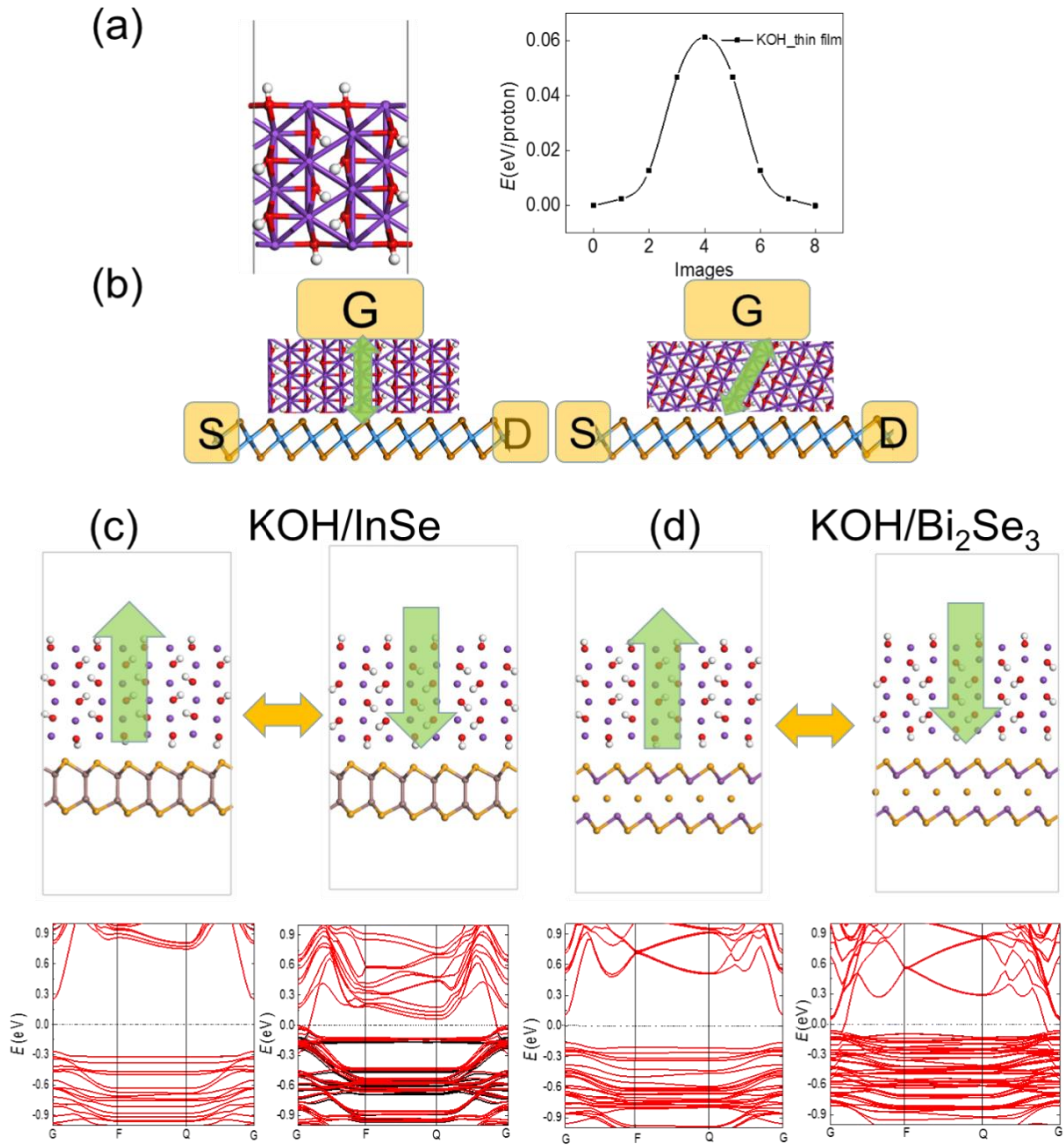

Fig. 4 (a) KOH thin-layer and vertical FE switching pathway. (b)Designs of 2D FE field-effect transistors based on (c) KOH/InSe and (d) KOH/$Bi_2Se_3$ heterostructures. Green arrows denote the polarization direction of KOH thin-film.

**Conclusion**

In summary, we show the first-principles evidence of robust proton-transfer FE with considerable polarization in NaOH and KOH, which enable commercial mass production. Moreover, even a small amount of proton vacancies can change the mode of proton-transfer from intra-unitcell to trans-unitcell, leading to a greatly enhanced polarization due to the prolonged transfer distance: e.g., the polarization of a $NaOH_{0.999}$ crystal with size ~1 μm can be ~300 μC/cm$^2$. A new type of multiferroic soliton with "mobile" magnetic moment can also be formed by proton vacancy, rendering a coupling between FE and magnetism for efficient data reading/writing. The NaOH and KOH thin-film possess vertical polarizations that are robust against depolarizing field, which can be used as FE field-transistors for nonvolatile control of insulating/metallic state for 2D materials and hold great promise as multifunctional devices. We note that both NaOH and KOH are hygroscopic, and the polarization will be much reduced when water is adsorbed into the crystal due to disorder. Chemically inert materials like boron nitride or silicon dioxides can be nice choices for protective layer against water. Since sodium and potassium hydroxides are common base in laboratory, we expect our predictions can advance experimental verifications of their proton-transfer FE and related future applications in nanoelectronics.

**Acknowledgements**

We are supported by the National Natural Science Foundation of China (Nos. 21573084). We thank Prof. Junming Liu for helpful discussions, and Shanghai Supercomputing Center for providing computational resources.

**Declaration of Interests**

The authors declare no competing interests.

**Supporting Information**

The Supporting Information is available free of charge on the website:

Bandstructures of NaOH and KOH, and molecular dynamics simulation of NaOH at 300K.

**References**

1. Dawber, M.; Rabe, K. M.; Scott, J. F. Physics of Thin-film Ferroelectric Oxides. *Rev. Mod. Phys.* **2005,** *77* (4), 1083-1130.

2. Hong, S.; Auciello, O.; Wouters, D. *Emerging Non-Volatile Memories*. Springer US: Boston, MA, 2014.

3. Junquera, J.; Ghosez, P. Critical Thickness for Ferroelectricity in Perovskite Ultrathin Films. *Nature* **2003,** *422* (6931), 506-509.

4. Fong, D. D.; Stephenson, G. B.; Streiffer, S. K.; Eastman, J. A.; Auciello, O.; Fuoss, P. H.; Thompson, C. Ferroelectricity in Ultrathin Perovskite Films. *Science* **2004,** *304* (5677), 1650-1653.

5. Wu, M.; Jena, J. The Rise of Two‐Dimensional Van Der Waals Ferroelectrics. *Wiley Interdiscip Rev Comput Mol Sci* **2018,** *8* (5), e1365.

6. Wu, M.; Burton, J. D.; Tsymbal, E. Y.; Zeng, X. C.; Jena, P. Hydroxyl-Decorated Graphene Systems as Candidates for Organic Metal-Free Ferroelectrics, Multiferroics, and High-Performance Proton Battery Cathode Materials. *Phys.Rev.B* **2013,** *87* (8), 081406.

7. Wu, M.; Burton, J. D.; Tsymbal, E. Y.; Zeng, X. C.; Jena, P. Multiferroic Materials Based on Organic Transition-Metal Molecular Nanowires. *J. Am. Chem. Soc.* **2012,** *134* (35), 14423-14429.

8. Tu, Z.; Wu, M.; Zeng, X. C. Two-Dimensional Metal-Free Organic Multiferroic Material for Design of Multifunctional Integrated Circuits. *J Phys Chem Lett* **2017,** *8* (9), 1973-1978.

9. Wu, M.; Duan, T.; Lu, C.; Fu, H.; Dong, S.; Liu, J. Proton Transfer Ferroelectricity/Multiferroicity in Rutile Oxyhydroxides. *Nanoscale* **2018,** *10* (20), 9509-9515.

10. Horiuchi, S.; Kumai, R.; Tokura, Y. Hydrogen-Bonding Molecular Chains for High-Temperature Ferroelectricity. *Adv. Mater.* **2011,** *23* (18), 2098-2103.

11. Kresse, G.; Furthmüller, J. Efficient Iterative Schemes for Ab Initio Total-Energy Calculations Using a Plane-Wave Basis Set. *Phys.Rev.B* **1996,** *54* (16), 11169-11186.

12. Kresse, G.; Furthmüller, J. Efficiency of Ab-Initio Total Energy Calculations for Metals and Semiconductors Using a Plane-Wave Basis Set. *Comp. Mater. Sci.* **1996,** *6* (1), 15-50.

13. Perdew, J. P.; Burke, K.; Ernzerhof, M. Generalized Gradient Approximation Made Simple. *Phys. Rev. Lett.* **1996,** *77* (18), 3865-3868.

14. Blöchl, P. E. Projector Augmented-Wave Method. *Phys.Rev.B* **1994,** *50* (24), 17953-17979.

15. Grimme, S. Semiempirical GGA-type Density Functional Constructed with a Long-Range Dispersion Correction. *J. Comp. Chem.* **2006,** *27* (15), 1787-1799.

16. Monkhorst, H. J.; Pack, J. D. Special Points for Brillouin-Zone Integrations. *Phys.Rev.B* **1976,** *13* (12), 5188-5192.

17. King-Smith, R. D.; Vanderbilt, D. Theory of Polarization of Crystalline Solids. *Phys.Rev.B* **1993,** *47* (3), 1651-1654.

18. Henkelman, G.; Uberuaga, B. P.; Jónsson, H. A Climbing Image Nudged Elastic Band Method for Finding Saddle Points and Minimum Energy Paths. *J.Chem.Phys.* **2000,** *113* (22), 9901-9904.

19. Baroni, S.; de Gironcoli, S.; Dal Corso, A.; Giannozzi, P. Phonons and Related Crystal Properties from Density-Functional Perturbation Theory. *Rev. Mod. Phys.* **2001,** *73* (2), 515-562.

20. Busing, W. R. Infrared Spectra and Structure of NaOH and NaOD. *J.Chem.Phys.* **1955,** *23* (5), 933-936.

21. Amm, D. T.; Segel, S. L.; Heyding, R. D.; Hunter, B. K. Proton Nuclear Magnetic Resonance and Line Shape Analysis in The Alkali Metal Hydroxides: LiOH, NaOH, KOH, and RbOH. *J.Chem.Phys.* **1985,** *82* (6), 2529-2534.

22. Kashimori, Y.; Kikuchi, T.; Nishimoto, K. The Solitonic Mechanism for Proton Transport in a Hydrogen Bonded Chain. *J.Chem.Phys.* **1982,** *77* (4), 1904-1907.

23. Stamenković, S.; Žakula, R. B. Solitons in Pseudo One-Dimensional Hydrogen Bonded Ferroelectrics. *Physica A* **1980,** *102* (3), 554-560.

24. Li, L.; Wu, M. Binary Compound Bilayer and Multilayer with Vertical Polarizations: Two-Dimensional Ferroelectrics, Multiferroics, and Nanogenerators. *ACS nano* **2017,** *11* (6), 6382-6388.

25. Yang, Q.; Xiong, W.; Zhu, L.; Gao, G.; Wu, M. Chemically Functionalized Phosphorene: Two-Dimensional Multiferroics with Vertical Polarization and Mobile Magnetism. *J. Am. Chem. Soc.* **2017,** *139* (33), 11506-11512.

26. Wu, M.; Dong, S.; Yao, K.; Liu, J.; Zeng, X. C. Ferroelectricity in Covalently functionalized Two-dimensional Materials: Integration of High-mobility Semiconductors and Nonvolatile Memory. *Nano Lett* **2016,** *16* (11), 7309-7315.

27. Bandurin, D. A.; Tyurnina, A. V.; Yu, G. L.; Mishchenko, A.; Zólyomi, V.; Morozov, S. V.; Kumar, R. K.; Gorbachev, R. V.; Kudrynskyi, Z. R.; Pezzini, S.; Kovalyuk, Z. D.; Zeitler, U.; Novoselov, K. S.; Patanè, A.; Eaves, L.; Grigorieva, I. V.; Fal'ko, V. I.; Geim, A. K.; Cao, Y. High Electron Mobility, Quantum Hall Effect and Anomalous Optical Response in Atomically Thin InSe. *Nat. Nano.* **2017,** *12* (3), 223-227.

28. Hong, S.; Nakhmanson, S. M.; Fong, D. D. Screening Mechanisms at Polar Oxide Heterointerfaces. *Rep Prog Phys* **2016,** *79* (7), 076501.

Table of Contents

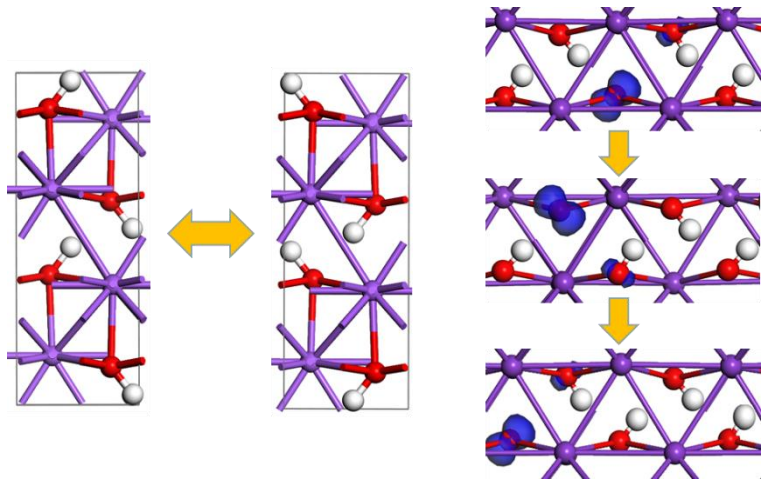